\documentclass[twocolumn,aps,floatfix]{revtex4}
\usepackage{amssymb} \usepackage{graphicx} \usepackage{amsmath}
\usepackage[T1]{fontenc} \usepackage{pstricks} \usepackage{subfigure}
\usepackage[colorlinks=true,citecolor=blue,urlcolor=red]{hyperref}
\usepackage[normalem]{ulem}
\usepackage{comment}

\begin{document}
\title{Three-dimensional Bose-Fermi droplets at nonzero temperatures}
\author{Maciej Lewkowicz,$\,^1$ Miros{\l}aw Brewczyk,$\,^2$ Mariusz Gajda,$\,^3$ and Tomasz Karpiuk$\,^2$ }
\affiliation{\mbox{$^1$ Doctoral School of Exact and Natural Sciences, University of Bia{\l}ystok, ul. K. Cio{\l}kowskiego 1K, 15-245 Białystok, Poland}
\mbox{$^2$ Wydzia{\l} Fizyki, Uniwersytet w Bia{\l}ymstoku,  ul. K. Cio{\l}kowskiego 1L, 15-245 Bia{\l}ystok, Poland}
\mbox{$^3$ Institute of Physics, Polish Academy of Sciences, Aleja Lotnik{\'o}w 32/46, PL-02668 Warsaw, Poland} \\ }

\date{\today}

\begin{abstract}
Using numerical methods, we study the formation of self-bound quantum Bose-Fermi droplets at nonzero temperatures. We describe an attractive atomic Bose-Fermi mixture using quantum hydrodynamics enriched by beyond-mean-field corrections and thermal fluctuations, together with a simplified self-consistent Hartree-Fock model. With these models, we determine that low-temperature droplets with finite lifetimes can exist in free space when the attraction between bosons and fermions is sufficiently strong. Additionally, Bose-Fermi droplets at nonzero temperatures can exist in a box potential in equilibrium with bosonic and fermionic vapor. We discuss the properties of Bose-Fermi droplets at nonzero temperatures in terms of the initial condensate fraction, total atom number, and interspecies attraction strength.
\end{abstract}

\maketitle

\section {Introduction}
Since the first experimental achievement of quantum degeneracy in fermionic potassium and lithium gases \cite{DeMarco99,Truscott01,Schreck01,Granade02,Hadzibabic03}, interest in ultracold mixtures with a fermionic component has grown rapidly. Studies of many-body quantum phenomena, including equilibrium and nonequilibrium thermodynamic properties \cite{Navon10,Ku12,Cao11}, correlations \cite{Rom06,Greif13,Hart15}, strong interactions in lower dimensions \cite{Murthy19,Luick20}, and dipolar effects \cite{Lu12,Aikawa14,Frisch14}, soon followed.

A novel quantum mixture -- a self-bound system of ultracold atoms -- has been realized experimentally in recent years. The first quantum droplets were observed in a gas of dysprosium-164 atoms \cite{Schmitt16}. The interplay of short-range and dipolar forces turned out to be crucial for stabilizing such systems. Other dipolar droplets were created using erbium-166 atoms \cite{Chomaz16}. The existence of a different kind of self-bound object -- droplets in a two-component mixture of bosonic potassium-39 atoms -- was soon demonstrated \cite{Cabrera18,Semeghini18}. Quantum droplets in heteronuclear bosonic mixtures were also observed \cite{Errico19}. Recently, a new member of the quantum droplet family appeared: self-bound droplets in a condensate of strongly dipolar molecules \cite{Zhang26}.

Another class of self-bound system -- Bose-Fermi droplets -- has recently been introduced at the theoretical level \cite{Rakshit18,Rakshit18a}. Such systems may bring a variety of physical phenomena into play, including connections to polaron physics, Cooper pairing, and fermionic superfluidity. Analogies can also be drawn with astronomical objects such as white dwarfs, due to the stabilization mechanism involving Fermi pressure. The quantum aspects of the disruption of a white dwarf star by a black hole have recently been modeled \cite{Karpiuk21,Nikolajuk25}, in connection with observations of recurrent ultraluminous X-ray flares \cite{Sivakoff05,Irwin16}.

This paper aims to answer the question of whether three-dimensional Bose-Fermi droplets exist at nonzero temperatures. Similar studies of attractive binary Bose droplets have been reported using the exact path-integral Monte Carlo method (see Refs. \cite{Spada23} and \cite{Spada24}). Here we study a dynamics of a Bose-Fermi mixture prepared initially in a thermal state of densities and interaction strenght corresponding to a droplet state, after releasing from a harmonic trap. We  use a hydrodynamic approach (see Appendix~\ref{first}), which is quite universal. For instance, it  allows to analyze the breathing modes of a Bose-Einstein condensate immersed in a Fermi sea, as observed in experiments (see Refs. \cite{Lous18, Huang19} and our results in Ref. \cite{Grochowski20}).  At low temperatures, the hydrodynamic equations for such a mixture can be expressed in terms of a bosonic wave function and a fermionic pseudo-wave function (see Refs. \cite{Rakshit18, Karpiuk20}). At higher temperatures, the bosonic wave function naturally transitions to a classical field according to a well-developed technique used to study the Bose gas at nonzero temperatures (see, for example, the reviews \cite{review, Proukakis08, Blakie08}).

A simplified description of the Bose-Fermi mixture at thermal equilibrium can also be formulated (see Appendix B). This approach is based on semiclassical distribution functions for bosons and fermions at a given temperature, making it straightforward to determine the equilibrium temperature corresponding to the stationary solution of the hydrodynamic equations. Unlike the equilibrium formulation, the hydrodynamic equations themselves do not explicitly contain the temperature as a parameter, since they are applicable to the dynamics of the mixture both in and far from thermal equilibrium.

Using the hydrodynamic description (see Appendix~\ref{first}), we find that Bose-Fermi droplets in free space exist at finite temperatures. They cool down to very low temperature within a finite amount of time by emitting bosonic and fermionic atoms. In a box potential, Bose-Fermi droplets cool down to a nonzero temperature via a similar mechanism and coexist with bosonic and fermionic vapors. This last finding is confirmed by a simplified, equilibrium-based approach.

The paper is organized as follows. In Sec. \ref{secII}, we propose a method for studying the formation of nonzero temperature Bose-Fermi droplets. Specifically, we develop a numerical technique to measure the droplet's temperature.  Then, in Sec. \ref{BFT}, we discuss the evolution and properties of a nonzero temperature Bose-Fermi droplets. We consider two scenario: when the droplet is in free space and when it is confined in a box (i.e., immersed in bosonic and fermionic vapors). Depending on the number of atoms and/or the strength of interparticle attraction, the Bose-Fermi droplet will either survive or explode. If the droplet survives, then in the first scenario it will cool down and equilibrate at a very low temperature within a finite amount of time. In the second scenario, the droplet will equilibrate at a temperature lower than the initial temperature. We conclude in Sec. \ref{concl}.

\section {Preparation of the thermal state}
\label{secII}

To study Bose-Fermi droplets at nonzero temperatures, we apply the following procedure. First, we determine the bosonic and fermionic densities for a mixture confined in a spherically symmetric harmonic trap at zero temperature, for a given boson-fermion attraction strength. To do so, we solve the set of hydrodynamic equations (\ref{eqB}) and (\ref{eqF}) using the imaginary-time technique (see Refs.~\cite{Gawryluk17,Swislocki26} and Appendix~\ref{third}). Next, we slowly pump energy into the mixture by randomizing the amplitude and phase of the zero-temperature bosonic field at each point of a numerical grid and evolving Eqs.~(\ref{eqB}) and (\ref{eqF}) in real time. The strength of the randomization determines the amount of energy injected into the system. Since both components are coupled, the injected energy spreads throughout the binary mixture. Assuming the energy is injected sufficiently slowly, the final state remains close to equilibrium, which can be verified by the characteristic dynamics of the condensate fraction that, in the final stages, does not change but only fluctuates around a mean value. Thus, a Bose-Fermi mixture in a state close to equilibrium is obtained. The condensate fraction is then below $1$, although the exact temperature of the sample -- clearly nonzero -- remains unknown. The temperature, however, can be uniquely related to the condensate fraction via the thermometry procedure described below. Finally, we open the trap and observe the subsequent dynamics of the Bose-Fermi mixture. This part of the procedure could, in principle, be realized experimentally. Non-spreading bosonic and fermionic densities indicate the formation of a self-bound Bose-Fermi droplet. Depending on the boundary conditions, we can observe either the evolution of a droplet at nonzero temperature in free space or the equilibrium dynamics of a droplet immersed in a saturated vapor of bosonic and fermionic atoms.

\begin{figure}[hbt]
\includegraphics[width=7.0cm]{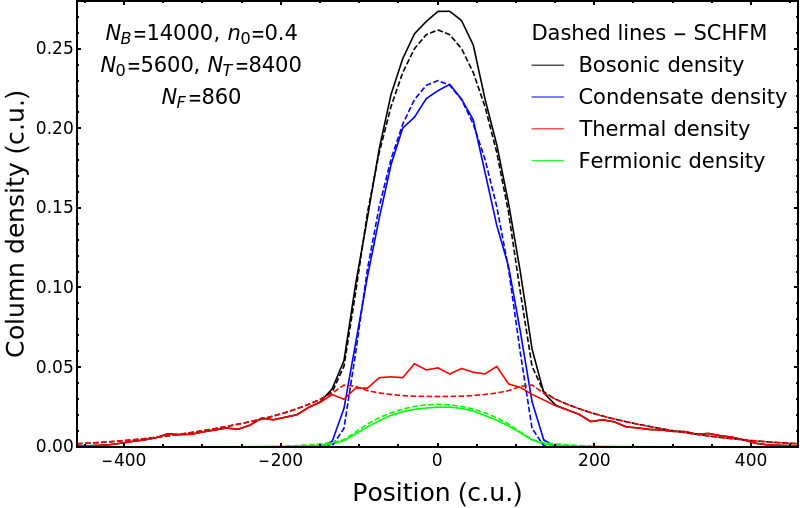} 
\caption{Column densities (in units of $1/a_B^3$) of a condensate and thermal fraction and of fermionic component for $a_{BF}/a_B=-3$ and the trapping frequencies $2\pi \times 250\,$Hz and $2\pi \times 4000\,$Hz for bosons and fermions, respectively. The number of condensate, thermal, and fermionic atoms is $N_0=5600$, $N_{th}=8400$, and $N_F=860$. The size of the clouds is measured in units of $a_B$ and $100$ code units equal about $1.3\,\mu$m (for $a_B=13.2\,$nm). The temperature of the sample is $T=268\,$nK which corresponds to the condensate fraction $n_0=0.4$. Solid lines show the results obtained using the hydrodynamic approach (see Appendix~\ref{first}) while the dashed lines comes from the simplified self-consistent Hartree-Fock method. }
\label{CFA_SCHFM_den}
\end{figure}

An example illustrating the outcome of the above prescription is shown in Fig.~\ref{CFA_SCHFM_den}, where the condensate, thermal, total, and fermionic column densities (i.e., densities integrated along one spatial dimension) are plotted in blue, red, black, and green solid lines, respectively. These curves are obtained within the hydrodynamic description for the interspecies scattering length $a_{BF}/a_B=-3$ and trapping frequencies of $2\pi \times 250\,$Hz for bosons and $2\pi \times 4000\,$Hz for fermions. We consider mixtures composed of $^{133}$Cs bosonic and $^{6}$Li fermionic atoms, which are currently studied experimentally \cite{Chin17,Chin18,Weidemuller14,Weidemuller25}. The numbers of bosons and fermions are specified in the figure caption.

After equilibration, we adopt a technique commonly used within the classical field approximation (CFA) to describe bosonic systems at nonzero temperatures \cite{review}. This standard procedure allows us to determine the condensate fraction, which is found to be $n_0=0.4$. We then apply thermometry based on the self-consistent Hartree-Fock method (SCHFM), as described in Appendix~\ref{second}. The corresponding densities are shown as dashed lines in Fig.~\ref{CFA_SCHFM_den}. They are obtained by matching the temperature so as to reproduce the same condensate fraction, $n_0=0.4$, which yields $T=268\,$nK, assuming $a_B=13.2\,$nm (see Appendix~\ref{third} for the code unit for temperature). 

\begin{figure}[hbt]
\includegraphics[width=7.0cm]{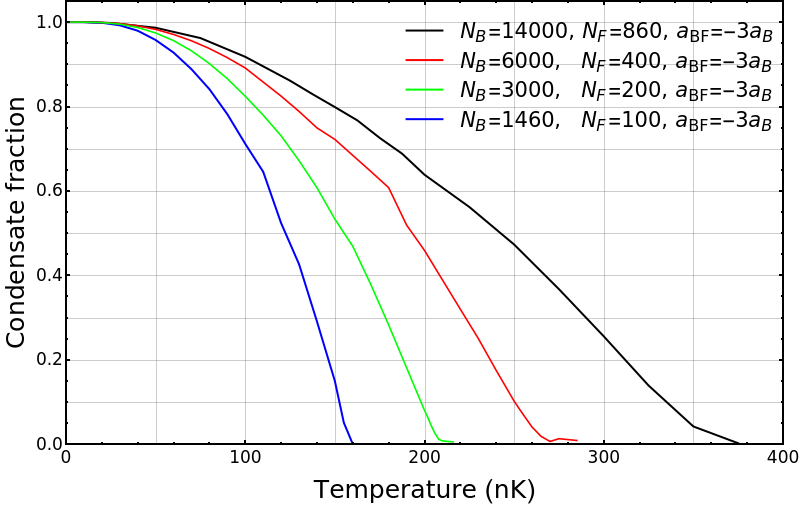}
\includegraphics[width=7.0cm]{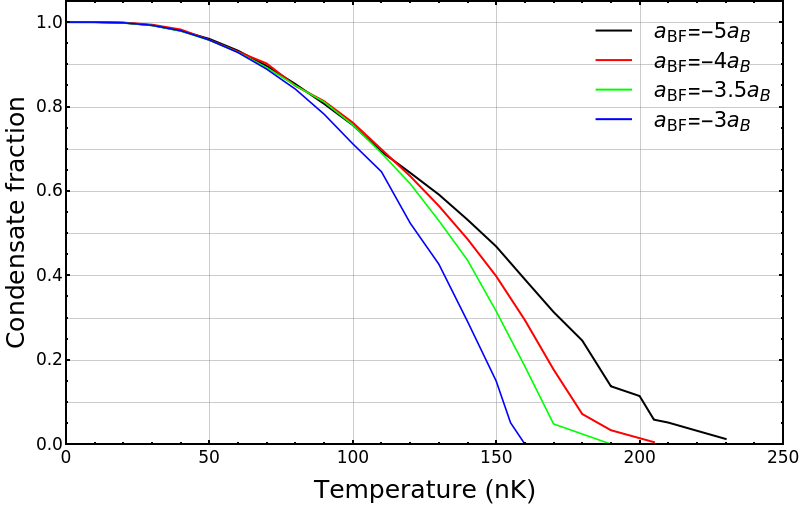}
\caption{Condensate fraction as a function of temperature. The top panel shows four samples distinguished by the number of atoms (given in the legend) and for $a_{BF}/a_B=-3$. The bottom panel shows four samples distinguished by the mutual scattering length (given in the legend) and for $N_B=1460$ and $N_F=100$. These curves were obtained by solving the self-consistent Hartree-Fock equations as described in Appendix~\ref{second}.}
\label{CFA_SCHFM_confrac}
\end{figure}

A summary of the thermometry procedure is presented in Fig.~\ref{CFA_SCHFM_confrac} for four sets of boson and fermion numbers, as indicated in the legend, with $a_{BF}/a_B=-3$ (top panel), and for four samples distinguished by their mutual scattering lengths (given in the legend) with $N_B=1460$ and $N_F=100$ (bottom panel). We would like to emphasize that the Hartree–Fock equilibrium model is used solely for thermometry, to characterize the initial state of the system. The SCHFM model enables us to determine the initial temperature of the sample prepared via CFA simulations. In the subsequent analysis of the Bose–Fermi droplet at nonzero temperatures, we use Eqs. (A1) and (A2). Note that Fig.~\ref{CFA_SCHFM_confrac} provides an estimation of the critical temperature for bosons based on classical field theory. Since bosons and fermions interact, this temperature differs from that obtained for an ideal Bose gas. For instance, an ideal sample of $N_B=14000$ particles in a spherically symmetric harmonic trap with a frequency of $2\pi\times 250\,$Hz (corresponding to the black curve in Fig.~\ref{CFA_SCHFM_confrac}) has a critical temperature of approximately $270\,$nK. For an interacting gas, however, the critical temperature is around $350\,$nK. Meanwhile, the Fermi temperature of the fermionic component (with $N_F=860$), assuming the ideal gas condition, increases tenfold to approximately $3300\,$nK.

\section{Bose-Fermi droplets at nonzero temperatures}
\label{BFT}

We now investigate the formation and properties of Bose-Fermi droplets at nonzero temperatures. The bosonic and fermionic densities presented in Fig.~\ref{CFA_SCHFM_den} are used as the initial conditions for the subsequent real-time evolution. Starting from these densities, we rapidly remove the trapping potential (within about $3.6\,$ms, approximately one oscillation period of the bosonic trap) and monitor the evolution of the mixture. We consider two scenarios determined by the applied boundary conditions. Absorbing boundary conditions correspond to a droplet (or Bose-Fermi mixture) evolving in free space, whereas periodic boundary conditions mimic confinement in a box potential. In the following, we analyze the properties of Bose-Fermi droplets with respect to the initial condensate fraction (i.e., temperature), the total atom number, and the strength of the interspecies attraction.

\subsection{Droplet evolution as a function of condensate fraction}

\begin{figure}[hbt]
\includegraphics[width=7.0cm]{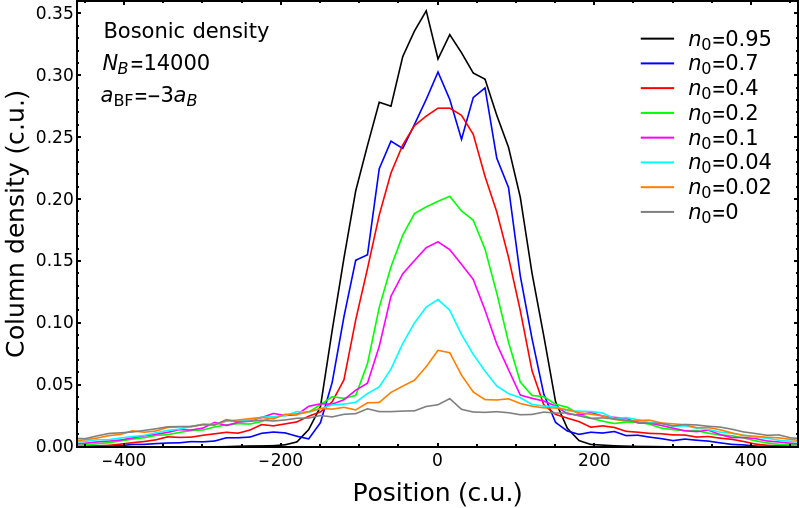} 
\caption{Column densities for the bosonic component are shown for $a_{BF}/a_B=-3$ and for different amounts of energy pumped into the system. The number of bosonic atoms is $N_B=14000$, and the condensate fraction is indicated in the legend. These results were obtained using the hydrodynamic approach. }
\label{nb14k_abs}
\end{figure}

Fig.~\ref{nb14k_abs} illustrates the bosonic column densities obtained using the hydrodynamic prescription for different amounts of energy injected into the system. The sample remains trapped, and the resulting equilibrium states correspond to a wide range of condensate fractions, from very small to large. The mixture contains $N_B=14000$ bosons and $N_F=860$ fermions, and the thermal fraction increases from top to bottom. The bosons and fermions interact attractively with $a_{BF}/a_B=-3$.

\begin{figure}[thb]
\includegraphics[width=7.0cm]{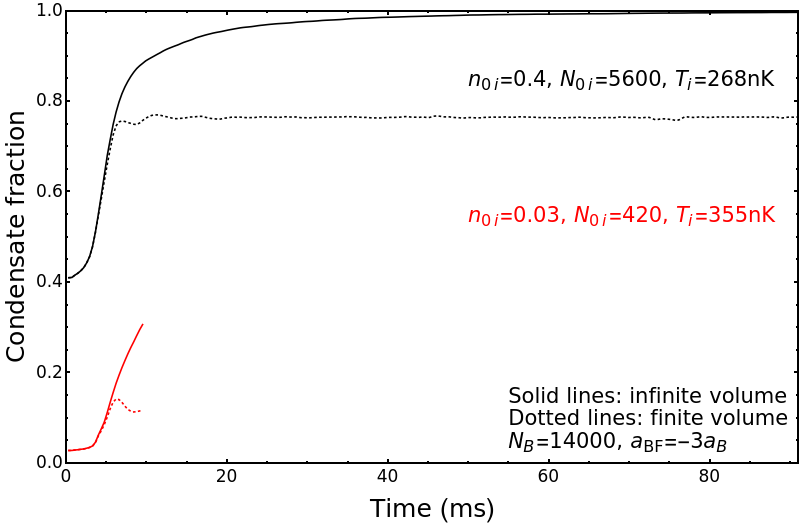} \\
\includegraphics[width=7.0cm]{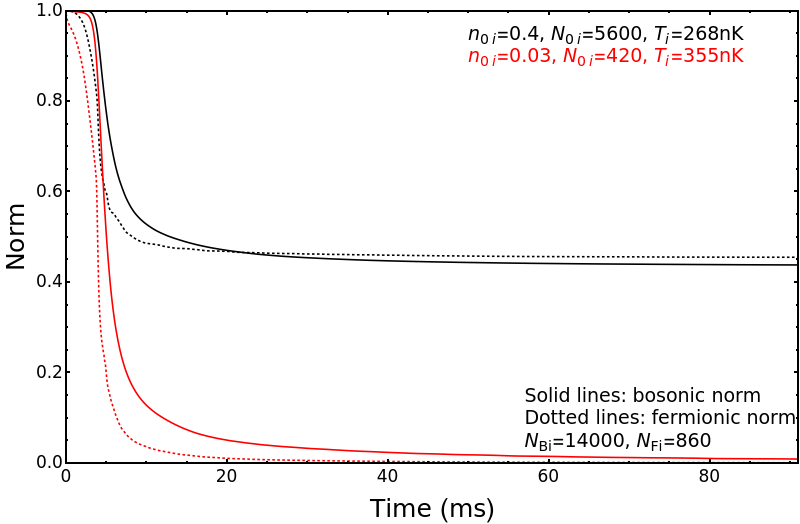}
\caption{Upper panel: Condensate fraction as a function of time (assuming $a_B=13.2\,$nm, see Appendix~\ref{third}) for the absorbing (solid line) and periodic (dotted line) boundary conditions.
Lower panel: Number of bosons (solid line) and fermions (dotted line) as a function of time for absorbing boundary conditions. The initial number of atoms is $N_B=14000$ and $N_F=860$, and $a_{BF}/a_B=-3$. The temperatures are $268\,$nK (black lines) and $355\,$nK (red lines), which corresponds to condensate fractions of $0.4$ and $0.03$, respectively. }
\label{confracnorm}
\end{figure}

Then, the trap is opened. Fig.~\ref{confracnorm} shows the condensate fraction (upper panel) and the bosonic and fermionic norms (lower panel) at later times. When the initial condensate fraction is set to $0.4$, the corresponding density is given by the red curve in Fig.~\ref{nb14k_abs}. Clearly, the droplet cools under both absorbing (solid lines) and periodic (dotted lines) boundary conditions, as evidenced by the increase of condensate fraction over time. Even though, in principle, the condensate fraction could increase at a constant temperature by adding atoms to the system, this is not the case here. Here, the number of atoms remains constant due to periodic boundary conditions or decreases due to absorbing boundary conditions (see the bottom panel of Fig.~\ref{confracnorm}). Indeed, the Bose-Fermi droplet cools down.

In the case of absorbing boundary conditions (i.e., a droplet evolving in free space), the condensate fraction approaches unity, indicating that the droplet temperature becomes very low, eventually approaching zero. Excess bosonic and fermionic atoms are removed through the boundaries, leading to a gradual decrease of their norms (see the lower panel of Fig.~\ref{confracnorm}). For periodic boundary conditions, corresponding to a droplet in a finite volume, the system also cools by expelling atoms into the surrounding vapor. This process eventually saturates, and the condensate fraction reaches a steady value below unity. The final state is a Bose-Fermi droplet in equilibrium with bosonic and fermionic vapor. The total number of particles is conserved, so the norms of both components remain equal to one.

The movies available at Refs.~\cite{abc0.4} and \cite{pbc0.4} show the evolution of the bosonic and fermionic column densities after the trap is opened. For a droplet evolving in free space \cite{abc0.4}, a form of self-evaporative cooling takes place. The most energetic atoms are no longer confined once the trap is removed and escape from the mixture, while the remaining atoms equilibrate at a lower temperature. This process continues until the bosonic component becomes an almost pure condensate. The initial stages of the evolution are rather violent due to large fluctuations in the thermal component; however, at later times the droplet becomes quiet, with a nearly stationary surface.

In contrast, when the droplet is confined to a finite volume \cite{pbc0.4}, its dynamics remain violent even after equilibrium with the surrounding vapor is established. The droplet is continuously hit by thermal atoms and exhibits irregular motion resembling a form of Brownian motion.

The fate of a Bose-Fermi droplet depends on the initial condensate fraction. For a sufficiently small initial condensate fraction ($n_0<0.04$, with other parameters as in Fig.~\ref{confracnorm}), the number of atoms in the droplet falls below the critical value required for its stability as it cools. Consequently, the droplet rapidly expels all atoms and disappears. The red lines in Fig.~\ref{confracnorm} show the condensate fraction (upper panel) and the norm (lower panel) for an initial condensate fraction of $0.03$. Both norms clearly drop to zero very quickly.

As in the previous case, the condensate fraction grows initially as the droplet cools by ejecting the most energetic atoms. After approximately $6\,$ms, the number of atoms in the droplet drops below the critical number (see the phase diagrams in Sec.~\ref{phasediagrams}). Then, the droplet explodes. Simultaneously, the size of the bosonic condensate increases, exceeding the size of the grid used to calculate the condensate fraction (see Appendix~\ref{third}). Therefore, monitoring the condensate fraction for times longer than $10\,$ms is not meaningful. This is why the red curves in Fig.~\ref{confracnorm} (upper panel) disappear after $10\,$ms. A movie illustrating the droplet’s explosion after approximately $6\,$ms is available at Ref.~\cite{abc0.04}.

\subsection{Droplet evolution as a function of the total number of atoms and interspecies attraction}
\label{phasediagrams}

\begin{figure}[h!bt]
\includegraphics[width=8.0cm]{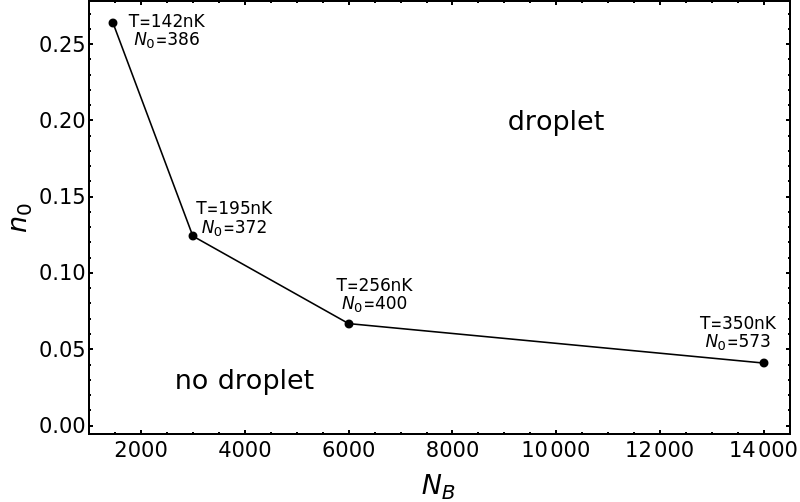} 
\caption{Phase diagram showing the region in which Bose-Fermi droplets exist, as a function of total number of bosons and the initial (i.e., before opening the trap) condensate fraction, for $a_{BF}/a_B=-3$. The number of fermions is $N_F=N_B/14.6$. Too high temperature (too low condensate fraction) destroys the droplet after the trap is removed. The data $(T,N_0)$ displayed in the diagram show the highest temperature at which the droplet is not destroyed and the corresponding number of condensed atoms for an initial total number of bosonic atoms $N_B$, as shown on the $x$-axis. }
\label{droplet_nodroplet_1}
\end{figure}

We now study how thermal Bose-Fermi droplets evolve as a function of the initial total number of atoms. In the phase diagram shown in Fig.~\ref{droplet_nodroplet_1}, the region above the black curve corresponds to conditions under which stationary droplets are formed when $a_{BF}/a_B=-3$ (as in Fig.~\ref{confracnorm}). The points on the curve indicate the critical initial condensate fraction required for droplet formation after the atoms are released from the trap. Periodic boundary conditions are used thereafter. Clearly, the system's fate depends on the initial number of atoms and the condensate fraction, which relates directly to the amount of thermal energy, i.e., temperature.

For a boson number of $N_B=1460$ (the leftmost point in Fig.~\ref{droplet_nodroplet_1}), the critical initial condensate fraction is larger than in the case studied in the previous subsection (corresponding to the rightmost point in Fig.~\ref{droplet_nodroplet_1}). As expected, for smaller atom numbers, thermal energy more easily drives evaporation and atom loss; hence, stable droplets are obtained only for larger initial condensate fractions. 

\begin{figure}[hbt]
\includegraphics[width=8.0cm]{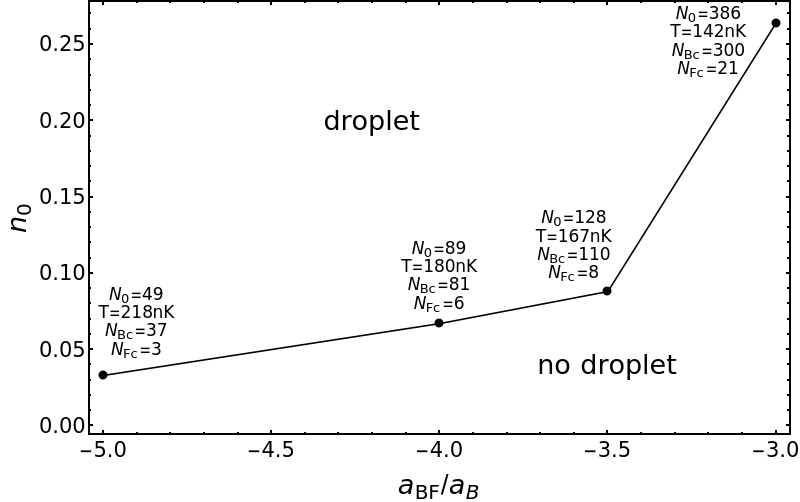} 
\caption{Phase diagram showing the region in which Bose-Fermi droplets exist as a function of interspecies attraction and the initial condensate fraction (i.e., the fraction of atoms in the condensate before opening the trap), for an initial number of bosons of $N_B=1460$. The displayed data $(T,N_0)$ show the maximum temperature at which the droplet survives opening the trap and the corresponding number of condensed atoms. The additional data, $(N_{Bc},N_{Fc})$, show the critical numbers of bosons and fermions, respectively, for a free droplet at zero temperature. This shows that the population of atomic droplets at the stability border decreases as boson-fermion attraction increases. A similar trend occurs with the ratio $N_B/N_F$. The same is true for infinitely large stable droplets (see Fig. 1 in Ref. \cite{Rakshit18}). In this case, $N_B/N_F$ equals $14.7$, $11.8$, $9.0$, and $7.2$ for $a_{BF}/a_B$ equal to $-3$, $-5$, $-7$, and $-9$, respectively. A too-small number of atoms does not allow for the formation of a droplet, and when the temperature is nonzero, it destroys the droplet after the trap is removed.
}
\label{droplet_nodroplet}
\end{figure}

Further understanding of droplet evaporation can be gained by examining Fig.~\ref{droplet_nodroplet}, which shows the phase diagram plotted in terms of the strength of the components’ mutual attraction and the initial condensate fraction. The minimal numbers of bosons and fermions, $N_{Bc}$ and $N_{Fc}$, respectively, required to support a stable free droplet at zero temperature are indicated in the figure.

When a droplet initially at nonzero temperature is released from the trap, it expels atoms on its way toward an  equilibrium. These expelled atoms are removed from the system through absorbing boundary conditions, leading to a gradual decrease in the total number of bosons and fermions. At some point, the number of bosons may drop below the black line in Fig.~\ref{droplet_nodroplet}, causing the droplet to disappear by rapidly ejecting the remaining atoms. Note that the black line closely matches the zero-temperature droplet data in free space in terms of the number of bosonic and fermionic atoms. For $a_{BF}/a_B=-3$ and an initial number of bosons $N_B=1460$, a droplet with as few as $400$ condensed bosons remains stable (see also Fig.~\ref{droplet_nodroplet_1}). Initial temperature of this droplet equals approximately $140\,$nK.

To stabilize even smaller droplets, the attraction between bosons and fermions must be stronger (see Fig.~\ref{droplet_nodroplet}). For instance, with an $a_{BF}/a_B$ ratio of $-5$, a Bose-Fermi droplet containing an initial number of condensed bosonic atoms equal to $N_0=50$ and at a temperature of approximately $220\,$nK can survive the opening of the trap (see the far-left point in Fig.~\ref {droplet_nodroplet}). However, such small droplets are highly sensitive to disturbances from thermal atoms during cooling and can be even shifted to random positions; see the movie at Ref.~\cite{abc0.05}.

\section{Conclusions}
\label{concl}
In summary, we have studied the formation of Bose-Fermi droplets at nonzero temperatures using a hydrodynamic description. These droplets exist when the boson-fermion attraction is sufficiently strong. We demonstrate that their lifetime is limited and that they cool by emitting bosonic and fermionic particles. In free space, they eventually reach very low temperature and thus act as nearly perfect coolers (see Fig.~\ref{confracnorm}). When confined in a box, the droplets reach equilibrium with the surrounding saturated bosonic and fermionic vapor (see Fig.~\ref{confracnorm}). Whether a droplet survives or explodes depends on the initial number of atoms, temperature, and a boson-fermion attraction strength (see Figs.~\ref{droplet_nodroplet_1} and \ref{droplet_nodroplet}).

\acknowledgments  
Part of the results were obtained using computers at the Computer Center of University of Bialystok.

\appendix
\section{Hydrodynamic equations}
\label{first}

We use the hydrodynamic approach \cite{Madelung,Frolich,Wong,MarchDeb,Grochowski17,Grochowski20,Karpiuk20,Ryszkiewicz22} to study the formation of atomic Bose-Fermi droplets. For this analysis, we start by describing the quantum mixture using the density and velocity fields of the bosonic and fermionic components. The basic equations for such fluids are the continuity equation and the Euler equation for each component. The fluids' motion is ideal, and the right-hand side of the Euler equations can be written as the gradient of the sum of different terms. These terms include quantum pressure for bosons, kinetic energy for fermions due to the Pauli exclusion principle, and energies resulting from external and interaction forces. We also consider the lowest-order gradient correction, known as the Weizs\"acker correction, to the kinetic energy of a fermionic gas. See Appendix B in Ref.~\cite{Rakshit18a} for a derivation of the hydrodynamic equations for a Bose-Fermi mixture.

We further assume that the fermionic flow is irrotational. From a numerical perspective, it is then convenient to recast the hydrodynamic equations into a Schrödinger-like form using the inverse Madelung transformation [48]. This transformation replaces the density and velocity fields with a single complex-valued function. Now, the bosonic field, denoted by $\psi_B({\bf{r}},t)$, and the fermionic pseudo-field, denoted by $\psi_F({\bf{r}},t)$, become the fundamental variables. The bosonic field, also known as the classical field, encompasses all the bosonic atoms within the system, including those in a condensate and those in a thermal cloud (see Ref. \cite{review}). The fermionic pseudo-field, specifically its square modulus and phase gradient, are interpreted as the density and velocity fields of a hydrodynamic fermionic fluid. The hydrodynamic equations of motion for the mixture are

\begin{eqnarray}
&&i\hbar \frac{\partial}{\partial t}  \psi_B = \left[-\frac{\hbar^2}{2 m_B} \nabla^2 
+ \frac{1}{2} m_B\, \omega_B^2\, r^2 + g_B n_B
\right.
\nonumber  \\
&& \left.  + g_{BF} n_F
+ \frac{5}{2} C_{LHY} n_B^{3/2}
+ C_{BF}\, n_F^{4/3}\,  A(w,\alpha)  \right.
\nonumber  \\
&&+ \left. 
C_{BF}\, n_B\, n_F^{4/3}\,  \frac{\partial A}{\partial \alpha} \frac{\partial \alpha}{\partial n_B}
\right]  \psi_B
\label{eqB}
\end{eqnarray}
and
\begin{eqnarray}
&&i\hbar \frac{\partial}{\partial t}  \psi_F = \left[-\frac{\hbar^2}{2 m_F} \nabla^2
+ (1-\xi) \frac{\hbar^2}{2 m_F} \frac{\nabla^2 |\psi_F|}{|\psi_F|} \right.
\nonumber  \\
&&+ \left. \frac{1}{2} m_F\, \omega_F^2\, r^2 + \frac{5}{3} \kappa\, n_F^{2/3}
+ g_{BF} n_B  \right.
\nonumber  \\
&&+ \left. \frac{4}{3} C_{BF}\, n_B\, n_F^{1/3}\,  A(w,\alpha) + 
C_{BF}\, n_B\, n_F^{4/3}\,  \frac{\partial A}{\partial \alpha} \frac{\partial \alpha}{\partial n_F}
\right]  \psi_F   \,.
\nonumber  \\
\label{eqF}
\end{eqnarray}

Note the presence of beyond mean-field corrections for boson-boson interactions (the term involving $C_{LHY}$ factor in Eq. (\ref{eqB})) and boson-fermion interactions (the last two terms in Eq. (\ref{eqB}) and Eq. (\ref{eqF})). Here, $g_B = 4\pi \hbar^2 a_B/m_B$ and $g_{BF} = 2\pi \hbar^2 a_{BF} /\mu$ are the coupling constants for boson-boson and boson-fermion interactions, respectively, with $a_B$ and $a_{BF}$ denoting the corresponding $s$-wave scattering lengths. The bosonic, fermionic, and the reduced masses are $m_B$, $m_F$, and $\mu$, respectively. Other coefficients are: $\kappa=(3/10)\, (6 \pi^2)^{2/3}\, \hbar^2/m_F$, $C_{LHY}=64/(15\sqrt{\pi})\,g_B\, a_B^{3/2}$, $C_{BF}=(6 \pi^2)^{2/3} \hbar^2 a_{BF}^2 / 2 m_F$. The coefficient $\xi$, which appears in the Weizs\"acker correction, equals $1/9$. The function $A(w,\alpha)$ has a form \cite{Viverit02}:
\begin{eqnarray}
&&A(w,\alpha) = \frac{2(1+w)}{3w}\left(\frac{6}{\pi}\right)^{2/3}\int^{\infty}_0 {\rm d}k \int^{+1}_{-1}{\rm d}{\Omega}
\nonumber\\
&&\left[1-\frac{3k^2(1+w)}{\sqrt{k^2+\alpha}}
\int^{1}_{0}\!{\rm d}q q^2 \frac{1-\Theta(1-\sqrt{q^2+k^2+2kq\Omega})}{\sqrt{k^2+\alpha}+wk+2qw\Omega}  \right], \nonumber  \\
\label{AAA}
\end{eqnarray}
where  $w=m_B/m_F$ and $\alpha = 16\pi\, n_B a_B^3 / (6\pi^2\, n_F a_B^3)^{2/3}$ are the dimensionless parameters and $\Theta()$ is the Heaviside step function. Spherically symmetric trapping is assumed with different frequencies for the bosonic and fermionic traps. The total densities of the bosons and fermions are calculated as $n_B=|\psi_B|^2$ and $n_F=|\psi_F|^2$, respectively. 

The set of Eqs. (\ref{eqB}) and (\ref{eqF}) was already used to observe the temperature effect on the breathing mode of a Bose-Einstein condensate immersed in a Fermi sea \cite{Grochowski20}. At very low temperatures, these equations enabled the study of the static and dynamic properties of self-bound quantum Bose-Fermi droplets \cite{Rakshit18,Karpiuk20}.

We note that, like any other mean-field-type approach, the pseudo-field description for the fermionic component may be limited when applied to the smallest droplets containing only a few particles. To some extent, these limitations are inherent to the method. Within the pseudo-field approach, the underlying single-particle spectrum is not accessible. Consequently, effects associated with shell structure, such as the distinction between open- and closed-shell configurations, are averaged out and do not appear as discrete features. Similarly, particle-hole excitations are not explicitly represented within the pseudo-field framework. In small systems, the corresponding excitation energies may constitute a significant fraction of, or even exceed, the droplet's binding energy. In these cases, a microscopic treatment may predict the droplet's destabilization or evaporation, whereas the pseudo-field description may still indicate a bound state. For small droplets, one can also consider anisotropic orbital occupations, which may be viewed, in a broader sense, as the finite-system counterpart of Fermi-surface distortions, although the pseudo-field description smooths them out. However, we note that these limitations become progressively less important as the droplet size increases. For larger droplets, shell effects are weaker, single-particle level spacing decreases, and collective properties dominate. In this regime, the pseudo-field description becomes increasingly accurate and is expected to reliably capture the relevant equilibrium and nonequilibrium properties.

\section{Self-consistent Hartree-Fock method}
\label{second}

In this approximation, the equilibrium densities of bosons and fermions are self-consistently determined for a given temperature based on semi-classical distribution functions. The semi-classical distribution functions for bosons and fermions are
\begin{eqnarray}
f_{\bf{p}}^B({\bf{r}}) = \frac{1}{e^{[\varepsilon_{\bf{p}}^B({\bf{r}})-\mu_B]/k_B T} - 1} 
\label{distfunB}
\end{eqnarray}
and 
\begin{eqnarray}
f_{\bf{p}}^F({\bf{r}}) = \frac{1}{e^{[\varepsilon_{\bf{p}}^F({\bf{r}})-\mu_F]/k_B T} + 1}    \,,
\label{distfunF}
\end{eqnarray}
respectively. For bosons, the semi-classical single-particle excitation energies are given by
\begin{eqnarray}
\varepsilon_{\bf{p}}^B({\bf{r}}) &=& \frac{{\bf{p}}^2}{2m_B} + 2\, g_B\, \big[ n_0({\bf{r}}) + n_{th}({\bf{r}}) \big] + V_{tr}^B({\bf{r}})   \nonumber  \\
&+& \frac{\delta E_{LHY}}{\delta n_0} + \frac{\delta E_{BF}}{\delta n_0}  \,,
\label{spenergyB}
\end{eqnarray} 
where the Lee-Huang-Yang term, responsible for beyond mean-field effects within bosons, is given by $E_{LHY}= C_{LHY}\, \int d{\bf{r}}\, n_0^{5/2}({\bf{r}})$ \cite{Lee57} and the boson-fermion interaction energy, $E_{BF}$, is the sum of the mean-field term $g_{BF} \int d{\bf{r}}\, \big[ n_0({\bf{r}}) + n_{th}({\bf{r}}) \big]\,n_F({\bf{r}}) $ and the correction due to quantum fluctuations, $C_{BF} \int d{\bf{r}}\, n_0({\bf{r}})\, n_F^{4/3}({\bf{r}})\, A(w,\alpha)$ \cite{Viverit02}. $n_0({\bf{r}})$ and $n_{th}({\bf{r}})$ ($n_B({\bf{r}})=n_0({\bf{r}})+n_{th}({\bf{r}})$) are the condensate and thermal densities, respectively (see Ref. \cite{Pethick} for discussion of the semi-classical energies for nonzero temperature bosonic gas within the Hartree-Fock approximation), and $n_F({\bf{r}})$ is the fermionic density. The thermal density for bosons is given as
\begin{eqnarray}
n_{th}({\bf{r}}) = \frac{1}{\lambda_{B}^3}\; g_{3/2}\{ e^{\left[ \mu_B - V_{eff}^B({\bf{r}}) \right]/k_B T} \}   \,,
\label{nBth}
\end{eqnarray}
where $\lambda_{B}=(2\pi \hbar^2/ m_B k_B T)^{1/2}$ and 
\begin{eqnarray}
V_{eff}^B({\bf{r}}) &=& 2\, g_B\, \big[ n_0({\bf{r}}) + n_{th}({\bf{r}}) \big] + V_{tr}^B({\bf{r}})    \nonumber  \\
&+& \frac{\delta E_{LHY}}{\delta n_0} + \frac{\delta E_{BF}}{\delta n_0}     \,.
\label{Veff}
\end{eqnarray} 
Similarly, the single-particle excitation energies for fermions are
\begin{eqnarray}
\varepsilon_{\bf{p}}^F({\bf{r}}) = \frac{{\bf{p}}^2}{2m_F} + V_{tr}^F({\bf{r}}) + \frac{\delta E_{BF}}{\delta n_F}  
\label{spenergyF}
\end{eqnarray} 
and the fermionic density becomes
\begin{eqnarray}
n_{F}({\bf{r}}) = \frac{1}{\lambda_{F}^3}\; f_{3/2}\{ e^{\left[ \mu_F - V_{tr}^F({\bf{r}}) - \delta E_{BF}/\delta n_F \right]/k_B T} \}  \,,
\label{nFer}
\end{eqnarray}
where the polylogarithm function of order $3/2$, $g_{3/2}(.)$, and the Fermi-Dirac integral function, $f_{3/2}(.)$, are standard functions for bosonic and fermionic gases in three dimensions \cite{Huang}. The chemical potential $\mu_B$ is determined from the formula (see Ref. \cite{Pethick})
\begin{eqnarray}
\mu_B = g_B \left[ n_0({\bf{r}}) + 2\, n_{th}({\bf{r}}) \right] + V_{tr}^B({\bf{r}}) + \frac{\delta E_{LHY}}{\delta n_0} + \frac{\delta E_{BF}}{\delta n_0} \,, \nonumber \\  
\label{chempot}
\end{eqnarray} 
while $\mu_F$ is calculated from the normalization condition $N_F=\int d{\bf{r}}\, n_F({\bf{r}})$. 

Based on the above equations, we can iteratively find the condensate density, $n_0({\bf{r}})$, the thermal density, $n_{th}({\bf{r}})$, and the fermionic density, $n_F({\bf{r}})$, for a given temperature and a specific number of bosonic and fermionic atoms. This is done in the following steps. First, given the normalized densities, $n_0({\bf{r}}), n_{th}({\bf{r}}),$ and $n_F({\bf{r}})$, the bosonic chemical potential is found from Eq. ($\ref{chempot}$) by choosing ${\bf{r}} = 0$. Next, the new thermal density ($n_{th}({\bf{r}})$) is calculated based on Eq. ($\ref{nBth}$). Next, the new condensate density, $n_0({\bf{r}})$, is obtained from Eq. (\ref{chempot}). Finally, a new fermionic density is determined from Eq. (\ref{nFer}) together with the value of the fermionic chemical potential by a normalization procedure. The whole cycle is then repeated until all densities converge. We slightly modify the above prescription: to find the condensate density, we solve the following differential equation (see Ref. \cite{Pethick})

\begin{eqnarray}
i\hbar \frac{\partial}{\partial t}  \psi_0({\bf{r}}) &=& \left\{-\frac{\hbar^2}{2 m_B} \nabla^2 
+ g_B\, \big[ n_0({\bf{r}}) + 2\, n_{th}({\bf{r}}) \big]
\right.   \nonumber  \\
&+& \left. V_{tr}^B({\bf{r}}) + \frac{\delta E_{LHY}}{\delta n_0} + \frac{\delta E_{BF}}{\delta n_0}  \right\}  \psi_0({\bf{r}}) 
\nonumber \\
\label{coneq}
\end{eqnarray}
and calculate the condensate density as $n_0({\bf{r}})=|\psi_0({\bf{r}})|^2$.

\section{Numerical details}
\label{third}
We use the following code units in our numerical calculations: $m_B$ (mass), $a_B$ (length), $\hbar^2/m_B a_B^2$ (energy), and $m_B a_B^2 / \hbar$ (time). We consider a $^{133}$Cs-$^6$Li mixture with $a_B=13.2294$ nm and $m_B=2.20695\times 10^{-25}$ kg. These parameters yield the following values of the remaining code units in SI units: $m_B a_B^2 / \hbar = 0.366267\,\mu$s and $\hbar^2/m_B a_B^2 = 2.87924\times 10^{-28}$ J. Additionally, the code unit for temperature is $\hbar^2/(m_B a_B^2 k_B) = 20.8354\,\mu$K.

In all numerical calculations presented in this paper, both within the CFA and SCHFM approaches, we use a computational grid of $128\times128\times128$ points. The spatial grid spacing is the same in all three directions, $\Delta x=\Delta y=\Delta z=15$ in code units. Consequently, the computational domain extends from $-960$ to $945$ in code units in each spatial direction. In SI units, this corresponds to a spatial step of $0.198441\,\mu$m and a simulation box extending from $-12.7\,\mu$m to $12.501\,\mu$m along each axis.

We start the numerical procedure by preparing the ground state in a spherically symmetric harmonic trap. The trap frequencies are $\omega_B = 2\pi\times 250$ Hz and $\omega_F = 2\pi\times 4000$ Hz for bosons and fermions, respectively. These values were chosen such that the size of the droplet in the trap is comparable to that of the free-space droplet at zero temperature. The ground state is obtained using the imaginary-time propagation method (see Refs.~\cite{Gawryluk17,Swislocki26}). The time step and total propagation time are set to $\Delta t=0.1$ and $10000$, respectively, in code units.

In the next step, we gradually inject energy into the system by randomizing the bosonic wave function while keeping the harmonic trap. This is achieved by solving Eqs. (\ref{eqB}) and (\ref{eqF}) in real time. The time step $\Delta t$ used during the real-time evolution is the same as that used in the imaginary-time evolution. At each time step, the bosonic wave function is perturbed according to
\begin{eqnarray}
 \psi_B({\bf r}_i) \rightarrow \psi_B({\bf r}_i) &+&
 c \,\left[r_{Re}({\bf r}_i)\, Re(\psi_B({\bf r}_i)) \right. \nonumber \\
 &+& \quad \left. r_{Im}({\bf r}_i)\, Im(\psi_B({\bf r}_i)) \right]\,,
\end{eqnarray}
where $c=0.002$, and $r_{Re}$ and $r_{Im}$ are random numbers drawn independently for each grid point ${\bf r}_i$ from the interval $(-0.5,0.5)$. The value of the parameter $c$ is chosen to satisfy two competing requirements: the energy should be injected into the system as rapidly as possible while keeping the system sufficiently close to thermal equilibrium throughout the process.
Since this procedure does not conserve the norm of the wave function, the bosonic wave function is renormalized after each randomization step. During the evolution, we monitor the condensate fraction and stop the heating process once the desired value is reached. Finally, the trap is switched off over a period of $10000$ code units ($3.66$ ms), and the subsequent evolution of the droplet is monitored.


To extract the condensate fraction, we employ the CFA method. Details of the CFA approach can be found in Ref. \cite{review}. According to this formalism, the condensate fraction and the thermal cloud are obtained from the averaged one-particle density matrix. We use the following averaging procedure:
\begin{multline}
 \bar{\rho}(x,y,x^{\prime},y^{\prime},t_2) = \frac{1}{t_2-t_1}\int_{t_1}^{t_2} dt \\
 \frac{1}{N_B}\int dz\, \psi_B(x,y,z,t)\, \psi_B^{*}(x^{\prime},y^{\prime},z,t) \,.
 \label{averaging}
\end{multline}
As can be seen, the averaging is performed over one spatial direction, which mimics the experimental imaging procedure. In addition, we average over a time interval of $t_2-t_1 = 1000$ code units. In our case, this is $0.366\,$ms, which is small compared to the characteristic timescale for changes in the condensate fraction (see Fig.~\ref{confracnorm}). Solving the eigenvalue problem for this coarse-grained one-particle density matrix yields the condensate fraction, the condensate wave function, and the thermal density.

One technical aspect requires further explanation. After integrating out the $z$ direction, we restrict the calculation to a $64\times64$ region of grid points centered around the origin in the $x$ and $y$ directions. Although the full grid consists of $128\times128$ points, solving the eigenvalue problem for a $16384\times16384$ matrix would be computationally too demanding. Fortunately, before releasing the trap, the condensate wave function, which corresponds to the lowest-momentum mode, is localized near the center of the system and remains well contained within the selected region.
The thermal component, however, can extend beyond the rectangular area used to construct the coarse-grained one-particle density matrix. Therefore, the boson number $N_B$ appearing in Eq. (\ref{averaging}) must correspond to the number of bosons contained within the selected region rather than the total number of bosons in the system.


Our numerical time-evolution algorithm is based on the Fourier transform, which naturally imposes periodic boundary conditions. Consequently, without any additional modifications, the system would correspond to a finite simulation volume with periodic boundaries. To mimic an effectively infinite system, we introduce absorbing terms, $i\Gamma(r)\psi_B$ and $i\Gamma(r)\psi_F$, on the right-hand sides of Eqs. (\ref{eqB}) and (\ref{eqF}), respectively.
The function $\Gamma$ is defined as $\gamma (r-r_0)$ for $r>r_0$ and $0$ for $r\le r_0$. The parameters $\gamma$ and $r_0$ are set to $-0.001$ and $800$, respectively, in code units. The value of $\gamma$ was determined through numerical tests and chosen to maximize absorption while minimizing reflections from the absorbing region.

\end{document}